\pdfoutput=1
\documentclass[times]{speauth}
\usepackage{tikz}
\usetikzlibrary{chains,positioning}

\usepackage{todonotes}
\usepackage{algorithm}
\usepackage[noend]{algorithmic} 

\usepackage{numcompress}

\makeatletter
\def\ps@pprintTitle{%
  \let\@oddhead\@empty
  \let\@evenhead\@empty
  \let\@oddfoot\@empty
  \let\@evenfoot\@oddfoot
}
\makeatother
\graphicspath{{../figures/}{.}}
\usepackage{booktabs}
\usepackage{paralist}
\usepackage{siunitx} 
\usepackage{graphicx}
\usepackage{amsmath}
\usepackage{url}
\usepackage[T1]{fontenc}
\usepackage{times}
\usepackage{subfig}
\usepackage{mathtools}

\DeclarePairedDelimiter\ceil{\lceil}{\rceil}
\DeclarePairedDelimiter\floor{\lfloor}{\rfloor}
\usepackage[colorlinks=false]{hyperref}

\begin{document}

\newcommand{\Censeighten}{\textsc{Census1881}}
\newcommand{\CensInc}{\textsc{CensusIncome}}
\newcommand{\Wikileaks}{\textsc{Wikileaks}}
\newcommand{\Weather}{\textsc{Weather}}
\newcommand{\Censtwothousand}{\textsc{Census2000}}

\runningheads{S.~Chambi, D. Lemire, O.~Kaser, R.~Godin}{Better bitmap performance with Roaring bitmaps}

\title{Better bitmap performance with Roaring bitmaps}

\author{S.~Chambi\affil{1}, D. Lemire\affil{2}, O.~Kaser\affil{3}, R.~Godin\affil{1}}

\address{\affilnum{1}D\'epartement d'informatique, UQAM, Montreal, Qc, Canada\break
\affilnum{2}LICEF Research Center, TELUQ, Montreal, QC, Canada\break
\affilnum{3}Computer Science and Applied Statistics,
UNB Saint John, Saint John, NB, Canada
}

\cgsn{Natural Sciences and Engineering Research Council of Canada}{261437}
\corraddr{Daniel Lemire, LICEF Research Center, TELUQ, 
Universit\'e du Qu\'ebec, 
5800 Saint-Denis,
Office 1105,
Montreal (Quebec),
H2S 3L5 Canada. Email: lemire@gmail.com}

%
%

\begin{abstract}
Bitmap indexes are commonly used in databases and search engines. By exploiting bit-level parallelism, they can significantly accelerate 
queries. However, they can use much memory, and thus we might prefer compressed bitmap indexes. Following Oracle's lead, bitmaps are often compressed using run-length encoding (RLE). Building on prior work, we introduce the \emph{Roaring} compressed bitmap format:
it uses packed arrays for compression instead of RLE\@.  We compare it to two  high-performance RLE-based bitmap encoding techniques: WAH (Word Aligned Hybrid compression scheme) and Concise (Compressed `n' Composable Integer Set). On synthetic and real data, we find that Roaring bitmaps \begin{inparaenum}[(1)]  \item often compress significantly better (e.g., $2 \times$) and
\item  are faster than the compressed alternatives (up to $900\times$ faster for intersections).  \end{inparaenum}
Our results challenge the view that RLE-based bitmap compression is best.
\end{abstract}



\maketitle

\section{Introduction}

A bitmap (or bitset) is a binary array that  we can view as an efficient and compact representation of an integer set, $S$. Given a bitmap of $n$~bits, the $i^{\mathrm{th}}$  bit is set to one if the $i^{\mathrm{th}}$ integer in the range $[0, n-1]$  exists in the set. For example, the sets $\{3,4,7\}$ and $\{4,5,7\}$ might be stored in binary form as \texttt{10011000}
and \texttt{10110000}. We can
compute the union or the intersection  between
two such corresponding lists using bitwise operations (OR, AND) on the bitmaps (e.g., \texttt{10111000} and \texttt{10010000} in our case).  
Bitmaps are part of the Java platform  (\texttt{java.util.BitSet}).

When the cardinality of $S$ 
is relatively large compared to 
the universe size, $n$ 
(e.g., $|S| > n /64$ on 64-bit processors), bitmaps are often superior to other comparable data structures  such as arrays, hash sets or trees. However, on moderately low density bitmaps 
($n/10000< |S| < n /64$), compressed bitmaps 
such as Concise can be preferable~\cite{Colantonio:2010:CCN:1824821.1824857}.

Most of the recently proposed compressed bitmap formats are
derived
 from Oracle's BBC~\cite{874730} and use run-length encoding (RLE) for compression:
WAH~\cite{wu2008breaking},
Concise~\cite{Colantonio:2010:CCN:1824821.1824857},
EWAH~\cite{arxiv:0901.3751}, COMPAX~\cite{netfli},
VLC~\cite{Corrales:2011:VLC:2033546.2033586},
 VAL-WAH~\cite{guzuntunable}, etc.  Wu et al.'s WAH is 
 probably the best known.
 WAH divides a bitmap of $n$~bits into $\ceil*{\frac{n}{w-1}}$~words of $w-1$~bits, where $w$ is a convenient word length (e.g., $w=32$). WAH distinguishes between two types of words: words made of just $w-1$~ones (\texttt{11$\cdots$1}) or just $w-1$~zeros (\texttt{00$\cdots$0}), are \emph{fill words}, whereas words containing a mix of zeros and ones (e.g., \texttt{101110$\cdots$1}) are \emph{literal words}.
 Literal words are stored using $w$~bits: the  most significant bit is set to zero and the remaining bits store the heterogeneous $w-1$~bits. Sequences of homogeneous fill words (all ones or all zeros) are also stored using $w$~bits: the most significant bit is set to 1, the second most significant bit indicates the bit value of the homogeneous word sequence, while the remaining $w-2$~bits store the run length of the homogeneous word sequence.

 When compressing a sparse bitmap, e.g., corresponding to the set $\{0,2(w-1), 4(w-1),\ldots\}$,  WAH can use $2w$~bits per set bit. Concise reduces this memory usage by half~\cite{Colantonio:2010:CCN:1824821.1824857}. It uses a similar format except for coded fill words. Instead of storing the run length
$r$ using $w-2$~bits, Concise uses only $w-2- \ceil*{\log_2 (w)}$~bits, setting aside $\ceil*{\log_2 (w)}$~bits as \emph{position} bits. 
These  $\ceil*{\log_2 (w)}$~
{position} bits encode a number $p\in [0,w)$.  When  $p=0$, we decode $r+1$~fill words. When it is non-zero, we decode $r$~fill words preceded by a word that has its $(p-1)^{\mathrm{th}}$~bit flipped compared to the following fill words. 
Consider the case where $w=32$.
Concise can code the set $\{0,62, 124,\ldots\}$ using only 32~bits/integer, in contrast to WAH which requires 64~bits/integer.

Though they reduce memory usage, these formats derived from BBC have slow random access compared to an  uncompressed bitmap. That is, checking or changing the $i^{\mathrm{th}}$~bit value is an $O(n)$-time~operation. Thus, though they represent an integer set, we cannot quickly check whether an integer is in the set. This makes them unsuitable for some applications~\cite{Culpepper:2010:ESI:1877766.1877767}. Moreover, RLE formats have  a limited ability to quickly skip data. For example, suppose that we are computing the bitwise AND between two compressed bitmaps. If one bitmap has long runs of zeros, we might wish to skip over the corresponding words in the other bitmap. Without an auxiliary index, this might  be impossible with formats like WAH and Concise.

Instead of using RLE and sacrificing random access, we propose to partition the space $[0,n)$ into \emph{chunks} and to store  dense and sparse chunks  differently~\cite{KaserLemireIS2006}. 
On this basis, we introduce a new bitmap compression scheme called \emph{Roaring}. %
Roaring bitmaps store  32-bit integers in a compact and efficient two-level indexing data structure. 
Dense chunks are stored using bitmaps; sparse chunks use  packed arrays of 16-bit integers.
In our example ($\{0,62, 124,\ldots\}$), it would use only $\approx 16$~bits/integer, half of Concise's memory usage. Moreover, on  the synthetic-data test proposed by Colantonio and Di Pietro~\cite{Colantonio:2010:CCN:1824821.1824857}, it is  at least four~times faster than WAH and Concise. In some instances, it can be hundreds of times faster.

Our approach is reminiscent of O'Neil and O'Neil's RIDBit external-memory system.  RIDBit is a B-tree of  bitmaps, where  a list is used instead when a chunk's density is too small.
However RIDBit fared poorly compared to FastBit---a WAH-based system~\cite{4318091}: FastBit was up to $10\times$ faster. 
In contrast to the negative results of O'Neil et al., we find that Roaring bitmaps can be several times faster than WAH bitmaps for in-memory processing.  Thus one of our main contributions is to challenge the belief---expressed by authors such as by Colantonio and Di Pietro~\cite{Colantonio:2010:CCN:1824821.1824857}---that WAH bitmap compression is the most efficient alternative.

A key ingredient in the performance of Roaring bitmaps are the new bit-count processor instructions (such as \texttt{popcnt}) that became available on desktop processors more recently (2008). Previously, table lookups were often used instead 
in systems like RIDBit~\cite{Rinfret:2001:BIA:375663.375669}, but they can be several times slower.
 These new instructions allow us to quickly compute the density of new chunks, and to efficiently extract the location of the set bits from a bitmap. 
 
 To surpass RLE-based formats such as WAH and Concise, we 
 also rely on  several algorithmic strategies (see \S~\ref{sec:logicalOp}). For example, when intersecting two sparse chunks, we may use an approach based on binary search  instead of a  linear-time merge like RIDBit. 
 Also, when merging two chunks, we  predict whether the result is dense or sparse to minimize wasteful conversions.  
In contrast, 
 O'Neil et al.\ report that  RIDBit converts chunks after computing them~\cite{Rinfret:2001:BIA:375663.375669}.

\section{Roaring bitmap}\label{sec:roaring}

We partition the range of 32-bit indexes ($[0,n)$) into chunks of $2^{16}$~integers sharing the same 16~most significant digits. 
 We use specialized containers to store their 16~least significant bits.

When a chunk contains no more than 4096~integers, we use a sorted array of packed 16-bit integers. When there are more than 4096~integers, we use a $2^{16}$-bit bitmap. 
Thus, we have two types of containers: an
array container for \emph{sparse} chunks and a bitmap container for \emph{dense} chunks.
The 4096 threshold insures that at the level of the containers, each integer uses no more than 16~bits: we either use $2^{16}$~bits for more than 4096~integers, using less than 16~bits/integer, or else we use exactly 16~bits/integer.

The containers are stored in a dynamic array  with the shared 16~most-significant bits: this serves as a first-level index. The array keeps the containers sorted by the 16~most-significant bits. 
We expect this first-level index to be typically small: when $n=\num{1000000}$, it contains at most 16~entries. Thus it should often remain in the CPU cache. The containers themselves should never use much more than 8\,kB.

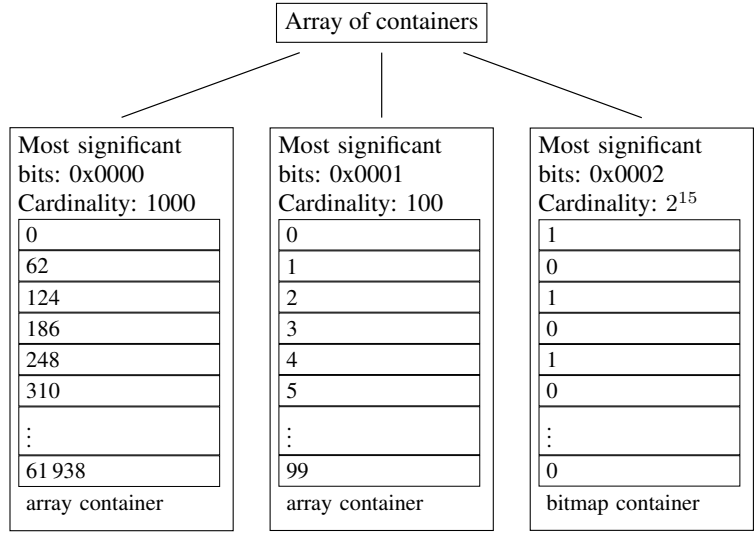
\begin{figure}\centering
\resizebox{0.7\columnwidth}{!}{\begin{tikzpicture}[edge from parent/.style={draw},
scale=0.9, every node/.style={scale=0.9},
start chain=1 going below,
start chain=2 going below,
start chain=3 going below,
start chain=4 going below,
start chain=5 going below,
start chain=6 going below,
start chain=10 going below,
start chain=7 going right,
start chain=8 going right,
start chain=9 going right,
start chain=11 going right,
child anchor=north,
level distance=4cm,
level 1/.style={sibling distance=34mm},
level 2/.style={sibling distance=10mm,level distance=3.5cm}]
\edef\sizetape{0.2cm}
\node{%
\begin{tikzpicture}[node distance=0cm]
    \tikzstyle{mytape}=[draw,minimum size=\sizetape]
    \node [on chain=9,mytape] {Array of containers};
\end{tikzpicture}
}
child{node {\begin{tikzpicture}[node distance=0cm]
    \tikzstyle{mytape}=[draw,minimum size=\sizetape]
    \node (id7) [on chain=7,mytape,text width=3cm] {Most significant bits: 0x0000\\Cardinality: 1000
    \begin{tikzpicture}[node distance=0cm]
    \tikzstyle{mytape}=[draw,minimum width=width("array container")]
    \node [on chain=1,mytape] {0};
    \node [on chain=1,mytape] {62};
    \node [on chain=1,mytape] {124};
    \node [on chain=1,mytape] {186} ;
    \node [on chain=1,mytape] {248};
    \node [on chain=1,mytape] {310};
    \node [on chain=1,mytape] {$\vdots$};
    \node(id1) [on chain=1,mytape] {61\,938};
      \node [below =of id1,node distance=0.7cm] {array container};
\end{tikzpicture}
    };
\end{tikzpicture}
}
}
child{node {\begin{tikzpicture}[node distance=0cm]
    \tikzstyle{mytape}=[draw,minimum size=\sizetape]
    \node (id7) [on chain=7,mytape,text width=3cm] {Most significant bits: 0x0001\\Cardinality: 100
    \begin{tikzpicture}[node distance=0cm]
    \tikzstyle{mytape}=[draw,minimum width=width("array container")]
    \node [on chain=1,mytape] {0};
    \node [on chain=1,mytape] {1};
    \node [on chain=1,mytape] {2};
    \node [on chain=1,mytape] {3} ;
    \node [on chain=1,mytape] {4};
    \node [on chain=1,mytape] {5};
    \node [on chain=1,mytape] {$\vdots$};
    \node(id1) [on chain=1,mytape] {99};
      \node [below =of id1,node distance=0.7cm] {array container};
\end{tikzpicture}
    };
\end{tikzpicture}
}
}
child{node {\begin{tikzpicture}[node distance=0cm]
    \tikzstyle{mytape}=[draw,minimum size=\sizetape]
    \node (id7) [on chain=7,mytape,text width=3cm] {Most significant bits: 0x0002\\Cardinality: $2^{15}$
    \begin{tikzpicture}[node distance=0cm]
    \tikzstyle{mytape}=[draw,minimum width=width("array container")]
    \node [on chain=1,mytape] {1};
    \node [on chain=1,mytape] {0};
    \node [on chain=1,mytape] {1};
    \node [on chain=1,mytape] {0} ;
    \node [on chain=1,mytape] {1};
    \node [on chain=1,mytape] {0};
    \node [on chain=1,mytape] {$\vdots$};
    \node(id1) [on chain=1,mytape] {0};
      \node [below =of id1,node distance=0.7cm] {bitmap container};
\end{tikzpicture}
    };
\end{tikzpicture}
}
};
\end{tikzpicture}}
\caption{Roaring bitmap containing the list of the first 1000~multiples of 62,   all integers 
$[2^{16},2^{16}+100)$  and  all even numbers in $[2\times 2^{16},3\times 2^{16})$.}
\label{fig:roaringcontainer}
\end{figure}

To illustrate the data structure, 
consider the list of the first 1000~multiples of 62,   all integers 
$[2^{16},2^{16}+100)$  and  all even numbers in $[2\times 2^{16},3\times 2^{16})$. When encoding this list using the Concise format, we use one 32-bit fill word for each of the 1000~multiples of 62, we use two additional fill words to include the list of numbers between  $2^{16}$ and $2^{16}+100$, and the even numbers in $[2\times 2^{16},3\times 2^{16})$ are stored as literal words.
In the Roaring format, both the multiples of~62 and the integers in $[2^{16},2^{16}+100)$ are stored using  an array container using 16-bit per integer. The even numbers in $[2\times 2^{16},3\times 2^{16})$ are stored in a $2^{16}$-bit bitmap container. See Fig.~\ref{fig:roaringcontainer}.

Each Roaring container keeps track of its cardinality (number of integers) using a counter. Thus computing the cardinality of a Roaring bitmap can be done quickly: it suffices to sum at most  $\ceil*{n/2^{16}}$~counters. It also makes it possible to support rank and select queries faster than with a typical bitmap: rank queries count the number of set bits in a range $[0,i]$ whereas select queries seek the location of the $i^{\mathrm{th}}$~set bit.

The overhead due to the containers and the dynamic array means that our memory usage can exceed 16~bits/integer. However, as long as the number of containers is small compared to the total number of integers, we should never use much more than 16~bits/integer. We assume that there are far fewer containers than integers. More precisely, we assume that the density typically exceeds \SI{0.1}{\percent} or that $n/|S|> 0.001$. When 
applications encounter integer sets with lower density (less than \SI{0.1}{\percent}), a bitmap is unlikely to be the proper data structure.

The presented Roaring data layout is intentionally simple. Several variations are possible. 
For very dense bitmaps, when there are more than $2^{16}-4096$ integers per container, we could store the locations of the zero bits  instead of a $2^{16}$-bit bitmap. Moreover, we could better compress sequences of consecutive integers. We leave the investigation of these possibilities as future work.

\section{Access  operations}

To check for the presence of a 32-bit integer $x$, we first seek
the container corresponding to $x/2^{16}$, using binary search. If a bitmap container is found, we access the $(x \bmod{2^{16}})^{\mathrm{th}}$~bit. If an array container is found, we use a binary search again. 

We insert and remove an integer $x$ similarly. We first seek the corresponding container. When the container found is a bitmap, we set the value of the corresponding bit and update the cardinality accordingly. If we find an array container, we use a binary search followed by a linear-time insertion or deletion. 

When removing an integer, a bitmap container might become an array container if its cardinality reaches 4096. When adding an integer, an array container might become a bitmap container when its cardinality exceeds 4096. When this happens, a new container is created with the updated data while the old container is discarded. Converting an array container to a bitmap container is done by creating a new bitmap container initialized with zeros, and setting the corresponding bits. To convert a bitmap container to an array container, we  extract the location of the set bits using an optimized algorithm (see Algorithm~\ref{algo:extract}).

\section{Logical operations}\label{sec:logicalOp}

We implemented various operations on Roaring bitmaps, including union (bitwise OR) and intersection (bitwise AND).
A bitwise operation between two Roaring bitmaps consists of iterating and comparing the 16~high-bits integers (keys) on the first-level indexes. For better performance, we maintain sorted first-level arrays. Two keys are compared at each iteration. On equality,  a second-level logical operation between the corresponding containers is performed. This always generates a new container. If the container is not empty,  it is added to the result along with the common key. Then iterators positioned over the first-level arrays  are incremented by one. When two keys are not equal,  the array containing the smallest one is incremented by one position, and if a union is performed, the lowest key and a copy of the corresponding container are added to the answer. When computing unions, we repeat until the two first-level arrays are exhausted. And when computing intersections, we terminate as soon as one array is exhausted.

Sorted first-level arrays allows  first-level comparisons in $O(n_1+n_2)$~time, where $n_1$ and $n_2$  are the respective lengths of the two compared arrays. 
We also maintain the array containers sorted for the same advantages. As containers can be represented with two different data structures, bitmaps and arrays, a logical union or intersection between two containers involves one of the three following scenarios: 
\begin{description}
\item [Bitmap vs Bitmap:] We iterate over 1024~64-bit words.
For unions, we  perform 1024~bitwise ORs and write the result to a new bitmap container. See Algorithm~\ref{algo:bitmapor}. The resulting cardinality is computed efficiently in Java using the \texttt{Long.bitCount} method.

\begin{algorithm}
\caption{\label{algo:bitmapor}Routine to compute the union  of two bitmap containers. 
 }
\centering
\begin{algorithmic}[1]
\STATE \textbf{input}: two bitmaps $A$ and $B$ indexed as arrays of 1024~64-bit integers 
\STATE \textbf{output}: a bitmap $C$ representing the union of $A$ and $B$, and its cardinality $c$
\STATE $c\leftarrow 0$
\STATE Let $C$ be indexed as an array of 1024~64-bit integers   
\FOR {$i \in \{1,2,\ldots, 1024\}$}
\STATE $ C_i \leftarrow A_i \mathrm{~OR~} B_i$
\STATE $ c \leftarrow c + \mathrm{bitCount}(C_i)$
\ENDFOR
\RETURN $C$ and $c$
\end{algorithmic}
\end{algorithm}

It might seem like computing bitwise ORs and computing the cardinality of the result would be significantly slower than merely computing the bitwise ORs.
However, four factors mitigate this potential problem. 

\begin{enumerate}

\item Popular
processors (Intel, AMD, ARM) have fast instructions to compute the number of ones in a word. Intel and AMD's \texttt{popcnt} instruction has a throughput as high as one operation per CPU cycle.
\item Recent Java implementations translate a call to \texttt{Long.bitCount} into such fast instructions.
\item Popular processors are superscalar: they can execute several operations at once. Thus, while we retrieve the next data elements, compute their bitwise OR and store it in memory, the processor can apply the \texttt{popcnt}  instruction on the last result and increment the cardinality counter accordingly. 
\item For inexpensive data processing operations, the processor may not run at full capacity  due to cache misses. 
\end{enumerate}
On the Java platform we used for our experiments, we estimate that we can compute and write bitwise ORs at 700\,million 64-bit words per second. If we further compute the cardinality of the result as we produce it, our estimated speed falls to about 500\,million words per second. That is, we suffer a speed penalty of about \SI{30}{\percent} because we maintain the cardinality.  In contrast, competing methods like 
 WAH and Concise must spend time to decode the word type before performing  a single bitwise operation. These checks may cause expensive branch mispredictions or impair superscalar execution.

For computing intersections, we use a less direct route.
First, we compute the cardinality of the result, using 1024~bitwise AND instructions. If the cardinality is larger than 4096, then we proceed as with the union, writing the result of bitwise ANDs to a new bitmap container. Otherwise, we create a new array container. We extract the set bits from the bitwise ANDs on the fly, using Algorithm~\ref{algo:extract}. See Algorithm~\ref{algo:bitmapand}.

\begin{algorithm}
\caption{\label{algo:extract}Optimized algorithm to convert the set bits in a bitmap into a list of integers. We assume two-complement's arithmetic. The function bitCount returns the Hamming weight of the integer.
 }
\centering
\begin{algorithmic}[1]
\STATE \textbf{input}:  an integer $w$ 
\STATE \textbf{output}: an array $S$ containing the indexes where a  1-bit can be found in $w$
\STATE Let $S$ be an initially empty list
\WHILE {$w\neq 0$}
\STATE $t \leftarrow w \textrm{~AND~} -w$ (cf. \cite[p.~12]{warr:hackers-delight-2e})
\STATE append bitCount($t-1$) to $S$
\STATE 
   $w \leftarrow w \textrm{~AND~} (w-1)$ (cf. \cite[p.~11]{warr:hackers-delight-2e})
\ENDWHILE
\RETURN $S$
\end{algorithmic}
\end{algorithm}

\begin{algorithm}
\caption{\label{algo:bitmapand}Routine to compute the intersection of two bitmap containers. The function bitCount returns the Hamming weight of the integer.
 }
\centering
\begin{algorithmic}[1]
\STATE \textbf{input}: two bitmaps $A$ and $B$ indexed as arrays of 1024~64-bit integers 
\STATE \textbf{output}: a bitmap $C$ representing the intersection of $A$ and $B$, and its cardinality $c$ if $c>4096$ or an equivalent array of integers otherwise 
\STATE $c\leftarrow 0$
\FOR {$i \in \{1,2,\ldots, 1024\}$}
\STATE $ c \leftarrow c + \mathrm{bitCount}(A_i \mathrm{~AND~} B_i)$
\ENDFOR
\IF{$c>4096$}
\STATE Let $C$ be indexed as an array of 1024~64-bit integers 
\FOR {$i \in \{1,2,\ldots, 1024\}$}
\STATE $ C_i \leftarrow A_i \mathrm{~AND~} B_i$
\ENDFOR
\RETURN $C$ and $c$
\ELSE
\STATE Let $D$ be an array of integers, initially empty 
\FOR {$i \in \{1,2,\ldots, 1024\}$}
\STATE append the set bits in $ A_i \mathrm{~AND~} B_i$ to $D$ using Algorithm~\ref{algo:extract}
\ENDFOR
\RETURN $D$
\ENDIF
\end{algorithmic}
\end{algorithm}

\item [Bitmap vs Array:] When one of the two containers is a bitmap and the other one is  a sorted dynamic array, the intersection can be computed very quickly: we iterate over the sorted dynamic array, and verify the existence of each 16-bit integer in the bitmap container. The result is written out to an array container. Unions are also efficient: we create a copy of the bitmap and simply iterate over the array, setting the corresponding bits. 

\item [Array vs Array:] 	For unions, if the sum of the cardinalities is no more than 4096, we use a merge algorithm between the two arrays. Otherwise, we set the bits corresponding to both arrays in a bitmap container. We then compute the cardinality using fast instructions. If the cardinality is no more than 4096, we convert the bitmap container to an array container (see Algorithm~\ref{algo:extract}). 

 For intersections, we use a simple merge (akin to what is done in merge sort) when the two arrays have cardinalities that differ by less than a factor of 64. 
  Otherwise, we use galloping intersections~\cite{Culpepper:2010:ESI:1877766.1877767}. 
  The result is always written to a new array container.
  Galloping is superior to a simple merge when one array ($r$) is much smaller than other one ($f$) because it can skip many comparisons. Starting from the beginning of both arrays, we pick the next available integer $r_i$ from the small array $r$ and seek an integer at least as large $f_j$ in the large array $f$, looking first at the next  value, then looking at a value twice as far, and so on.
Then, we use binary search to advance in the second list to the first value larger or equal to $r_i$.

\end{description}

We can also execute some of these operations \emph{in place}: 
\begin{itemize}
\item  When computing the  union between two bitmap containers, we can modify one of the two bitmap containers instead of generating a new bitmap container.
Similarly, for the intersection between two bitmap containers,  we can modify one of the two containers if the cardinality of the result exceeds 4096.

\item  When computing the union between an array and a bitmap container, we can write the result to the bitmap container, by iterating over the values of the array container and setting the corresponding bits in the bitmap container. We can update the cardinality each time by checking whether the word value has been modified.
\end{itemize}
In-place operations can be faster because they avoid allocating and initializing new memory areas.


When aggregating many bitmaps, we use other optimizations. For example, when computing the union of many bitmaps (e.g., hundreds), we first locate all containers having the same key (using a priority queue). If one such container is a bitmap container, then we can clone this bitmap container (if needed) and compute the union of this container with all corresponding containers in-place. In this instance, the computation of the cardinality can be done once at the end. See Algorithm~\ref{algo:horizontalor}.

\begin{algorithm}
\caption{\label{algo:horizontalor}Optimized algorithm to compute the union of many roaring bitmaps 
 }
\centering
\begin{algorithmic}[1]
\STATE \textbf{input}:  a set $R$ of Roaring bitmaps as collections of containers; each container has a cardinality and a 16-bit key
\STATE \textbf{output}: a new Roaring bitmap $T$ representing the union 
\STATE Let $T$ be an initially empty Roaring bitmap. 
\STATE Let $P$ be the min-heap of containers in the bitmaps of $R$, configured to order the containers by their 16-bit keys. 
\WHILE {$P$ is not empty}
\STATE Let $x$ be the root element of $P$. Remove from the min-heap $P$ all elements having the same key as $x$, and call the result $Q$. 
\STATE Sort $Q$ by descending cardinality; $Q_1$ has maximal cardinality. 
\STATE Clone $Q_1$ and call the result $A$. The container $A$ might be an array or bitmap container.
\FOR {$i \in \{2,\ldots, |Q|\}$}
\IF {$A$ is a bitmap container}
\STATE  Compute the in-place union of $A$ with $Q_i$: $A\leftarrow A \mathrm{~OR~} Q_i$. Do not re-compute the cardinality of $A$: just compute the bitwise-OR operations.
\ELSE
\STATE  Compute the union of the array container $A$ with the array container $Q_i$: $A\leftarrow A \mathrm{~OR~} Q_i$. 
If $A$ exceeds a cardinality of 4096, then it becomes a bitmap container.
\ENDIF
\ENDFOR
\STATE If $A$ is a bitmap container, update $A$ by computing its actual cardinality.
\STATE Add $A$ to the output of Roaring bitmap $T$.
\ENDWHILE
\RETURN $T$
\end{algorithmic}
\end{algorithm}

\begin{figure}[t!]\centering
\subfloat[Compression: uniform distribution\label{fig:cu}]{\includegraphics[width=0.5\textwidth]{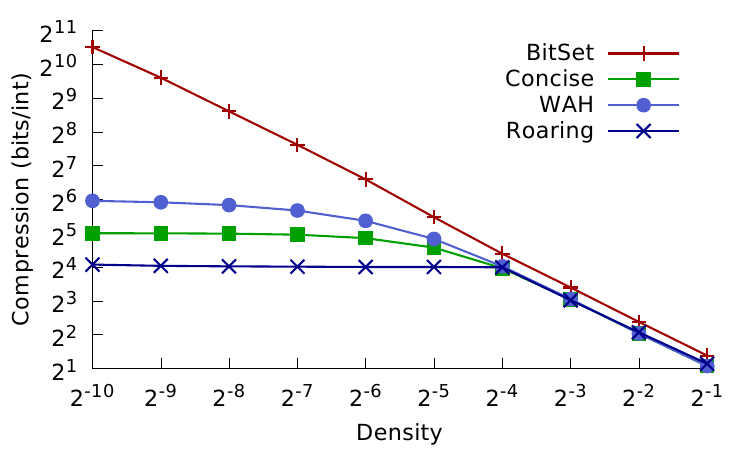}}
\subfloat[Compression: $\mathrm{Beta(0.5,1)}$  distribution\label{fig:cz}]{\includegraphics[width=0.5\textwidth]{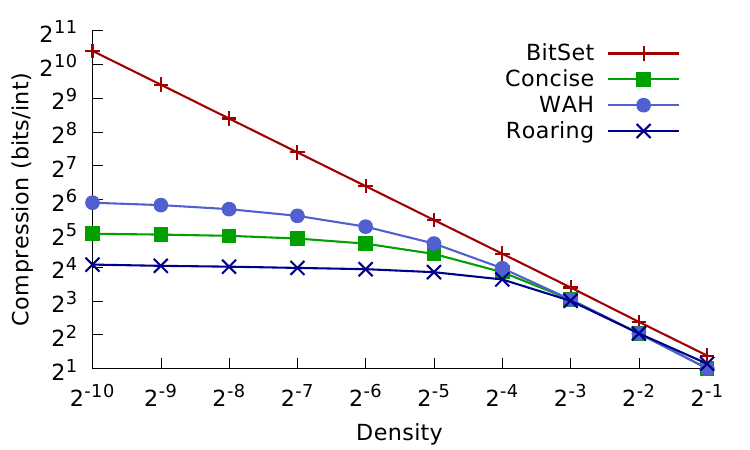}}

\subfloat[Intersection: discretized $\mathrm{Beta(0.5,1)}$ distribution\label{fig:iu}]{\includegraphics[width=0.5\textwidth]{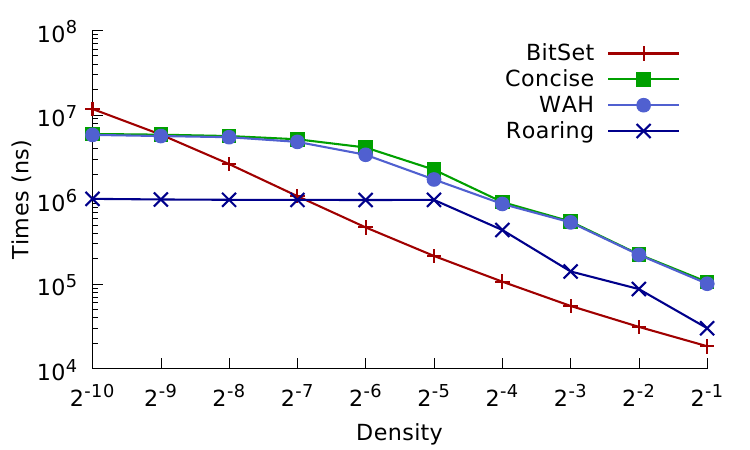}}
\subfloat[Union: discretized $\mathrm{Beta(0.5,1)}$ distribution\label{fig:iz}]{\includegraphics[width=0.5\textwidth]{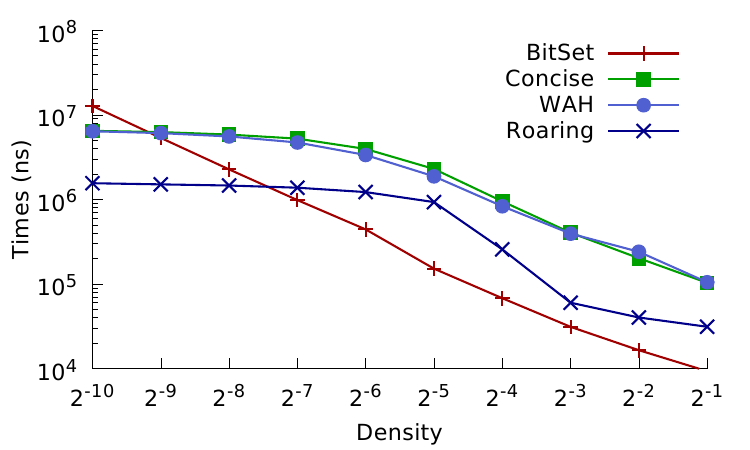}}

\subfloat[Append: uniform distribution\label{fig:au}]{\includegraphics[width=0.5\textwidth]{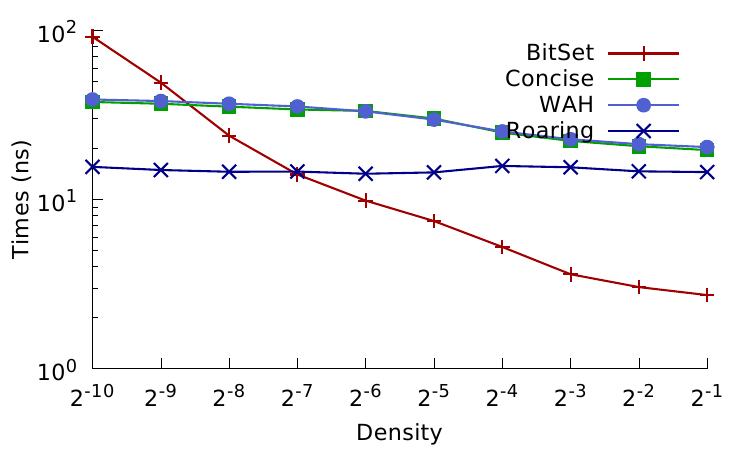}}
\subfloat[Removal: uniform distribution\label{fig:ru}]{\includegraphics[width=0.5\textwidth]{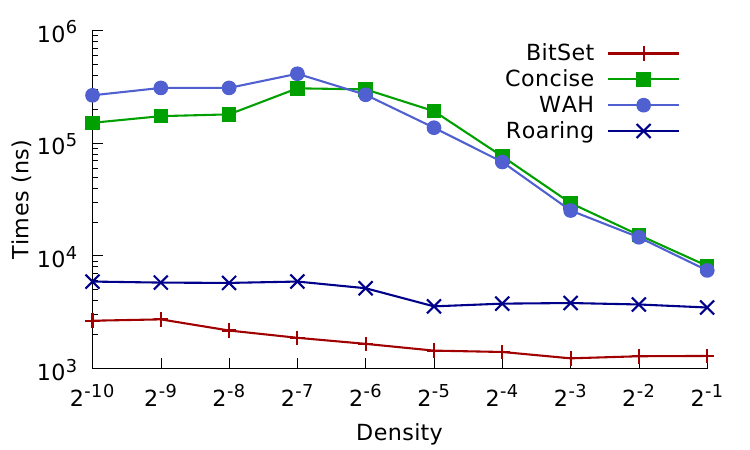}}
\caption{\label{fig:speed}Times and compression measurements: average of 100~runs}
\end{figure}

\section{Experiments}\label{sec:experiments}
We performed a series of experiments to compare the time-space performance of Roaring bitmaps with the performance of other well-known bitmap indexing schemes: Java's \texttt{BitSet}, WAH and Concise. We use the CONCISE Java library for WAH and Concise (version 2.2) made available by Colantonio and Di Pietro~\cite{Colantonio:2010:CCN:1824821.1824857}. 

Our Roaring-bitmap implementation code and data is freely available at \url{http://roaringbitmap.org/}. Our software has been thoroughly tested: our   Java library has been adopted by major database systems like Apache Spark~\cite{Zaharia:2010:SCC:1863103.1863113} 
and Druid~\cite{Yang:2014:DRA:2588555.2595631}. 

Benchmarks were performed on an AMD~FX\texttrademark{}-8150 eight-core processor running at 3.60\,GHz and having 16\,GB of RAM\@.  We used the Oracle Server JVM version~1.7 on Linux Ubuntu~12.04.1~LTS\@. All our experiments run entirely in memory.


To account for the just-in-time compiler in Java, we first run tests without recording the timings. Then we repeat the tests several times and report an average.

\subsection{Synthetic experiments}
We began by reproducing Colantonio and Di Pietro's synthetic experiments~\cite{Colantonio:2010:CCN:1824821.1824857}. However, while they included alternative data structures such as Java's \texttt{HashSet}, we focus solely on bitmap formats for simplicity.
Our results are generally consistent with Colantonio and Di Pietro's results, given the fact that we have a better processor.

Data sets of $10^{5}$ integers were generated according to two synthetic data distributions: uniform and discretized $\mathrm{Beta(0.5,1)}$ distributions. (Colantonio and Di Pietro described the latter as a \emph{Zipfian} distribution.) The four schemes were compared on several densities~$d$ varying from $2^{-10}$ to $0.5$. To generate an integer, we first picked a floating-point number $y$ pseudo-randomly in $[0, 1)$. When we desired a uniform distribution, we added $\floor*{y \times \textrm{max}}$  to the set. In the $\beta$-distributed case, we added  $\floor*{y^{2} \times \textrm{max}}$. 
 The value \textrm{max} represents the ratio between the total number of integers to generate and the desired density ($d$) of the set, i.e.: $\textrm{max} = 10^{5} / d$. Because results on uniform and $\mathrm{Beta(0.5,1)}$ distributions are often  similar, we do not systematically present both.

We stress that our data distributions and benchmark closely follow Colantonio and Di Pietro's work~\cite{Colantonio:2010:CCN:1824821.1824857}. Since they used this benchmark to show the superiority of Concise over WAH, we feel that it is fair to use their own benchmark to assess our own proposal against Concise.

Figs.~\ref{fig:cu} and~\ref{fig:cz} show the average number of bits used by Java's \texttt{BitSet}  and the three bitmap compression techniques to store an integer in a set. In these tests, Roaring bitmaps require \SI{50}{\percent}  of the space consumed by Concise and \SI{25}{\percent} of WAH space on sparse bitmaps.

The \texttt{BitSet} class uses slightly more memory even for dense bitmaps in our tests. This is due to its memory allocation strategy that doubles the size of the underlying array whenever more storage is required. We could recover this wasted memory by cloning the newly constructed bitmaps. Our roaring  bitmap implementation has a \texttt{trim} method that can be used to get the same result. We did not call these methods in these tests.

We also report on intersection and union times. That is, we take two bitmaps and generate a new bitmap representing the intersection or union. For the \texttt{BitSet}, it means that we first need to create a copy (using the  \texttt{clone} method), since bitwise operations are in-place.
Figs.~\ref{fig:iu} and~\ref{fig:iz} present the average time in  nanoseconds  to perform intersections and unions between two sets of integers. Roaring bitmaps are $\times4-\times5$~times faster than Concise and WAH for intersections on all tested densities. Results for unions are similar except that for moderate densities ($2^{-5}\leq  d\leq 2^{-4}$), Roaring is only moderately (\SI{30}{\percent}) faster than Concise and WAH. \texttt{BitSet} outperforms the other schemes on dense data, but it is $>10 \times$ slower on sparse bitmaps.

We measured times required by each scheme to add a single element $a$ to a sorted set $S$ of integers, i.e.: $\forall i \in S: a > i$. Fig.~\ref{fig:au} shows that Roaring requires less time  than WAH and Concise. Moreover, WAH and Concise do not support the efficient insertion of values in random order, unlike Roaring bitmaps.
Finally, we  measured the time needed to remove one randomly selected element from an integers set (Fig.~\ref{fig:ru}). We observe that Roaring bitmaps have much better results than the two other compressed formats.


\subsection{Real-data experiments}
Tables~\ref{tab:real-data-char}--\ref{tab:real-data}
present results for the five real data sets used an earlier study of
compressed bitmap indexes~\cite{LemireKaserGutarra-TODS}. 
There are only two  exceptions:
\begin{itemize}
\item We only use the September~1985 data for the \Weather{} data set
(an approach others have used before~\cite{304214}), which was otherwise too large for our test
environment. 
\item We omitted the \Censtwothousand{} data set because it contains only  bitmaps having an average cardinality of~30 over a
large universe ($n=\num{37019068}$). It is an ill-suited scenario for bitmaps. Because of the structure overhead, Roaring bitmaps used $4 \times$ as much memory as Concise bitmaps. Still, Roaring bitmaps were about $4 \times$ faster when computing intersections. 
\end{itemize}
The data sets were taken as-is: we did not sort them prior to indexing.

For each data set, a bitmap index was built.  Then we
chose 200~bitmaps from the index, using an approach similar to
stratified sampling 
 to control for the large range of attribute
cardinalities.  We first sampled 200~attributes, with replacement.  For
each sampled attribute, we selected one of its bitmaps uniformly at
random.  The 200~bitmaps were used as 100~pairs of inputs for 100~pairwise ANDs and ORs; 
Tables~\ref{tab:real-data-and-speed}--\ref{tab:real-data-or-speed}
show the time factor increase if Roaring is replaced by \texttt{BitSet},
WAH or Concise.  (Values below~1.0 indicate cases where Roaring 
was slower.) 
Table~\ref{tab:real-data-compression} shows the storage
factor increase when Roaring  is replaced by one of the other
approaches.

\begin{table}
\centering
\caption{\label{tab:real-data-char} Sampled bitmap characteristics and Roaring size. 
}
\begin{tabular}{crrrrr}\toprule
        & \Censeighten{} & \CensInc{} & \Wikileaks{} & \Weather{} 
        \\
        \midrule
Rows    & \num{4277807}     &    \num{199523}  &  \num{1178559}     &  \num{1015367}   
\\
Density & \num{1.2e-3}      &    \num{1.7e-1}  &  \num{1.3e-3}      &  \num{6.4e-2}   
\\
Bits/Item & \num{18.7}      &    \num{2.92}    &  \num{22.3}        &  \num{5.83}    
 \\
\bottomrule
\end{tabular}
\end{table}

 \begin{table}
\centering
\caption{Results on real data\label{tab:real-data}}
\subfloat[\label{tab:real-data-compression} Size expansion if Roaring is replaced with other schemes.]{%
\sisetup{%
round-mode=figures,round-precision=2
}
\begin{tabular}{cSSSS}
\toprule
        & \Censeighten{} & \CensInc{} & \Wikileaks{} & \Weather{} 
        \\
        \midrule
Concise &  2.21               &   \num{1.38}                &  0.79              &    1.38          
 \\
WAH     &  2.43               &    1.63                &  0.79              &    1.51          
         \\
BitSet  & 41.50                 &    2.89                &  55.45                &    3.49          
 \\\bottomrule
\end{tabular}}

\subfloat[\label{tab:real-data-and-speed} Time increase, for AND,  if Roaring is replaced with other schemes.]{%
\sisetup{%
round-mode=figures,round-precision=2
}
\begin{tabular}{cSSSS}\toprule
& \Censeighten{} & \CensInc{} & \Wikileaks{} & \Weather{} 
\\
\midrule
Concise & 921.81               &  6.58                   &    8.30              &   6.26       
   \\
WAH     &  841.08               &  5.89                  &    8.16              &     5.40          
 \\
BitSet  &  733.85               &  0.42                 &   27.91             &    0.64         
\\
\bottomrule
\end{tabular}}

\subfloat[\label{tab:real-data-or-speed} Time increases, for OR,  if Roaring is replaced with other schemes.]{%
\sisetup{%
round-mode=figures,round-precision=2
}
\begin{tabular}{cSSSS}\toprule
 & \Censeighten{} & \CensInc{} & \Wikileaks{} & \Weather{} 
\\
\midrule
Concise &   33.80               &  5.41                  &     2.14             &   3.87         
 \\
WAH     &   30.58               & 4.85                  &    2.06            &   3.39          
\\
BitSet  &   28.73               &  0.43                 &    6.72            &  0.48          
\\
\bottomrule
\end{tabular}
}
\end{table}

Roaring bitmaps are always faster, on average, than WAH and Concise.
On two data sets (\Censeighten{} and \Wikileaks{}), Roaring bitmaps are faster than \texttt{BitSet} while using much less memory ($40 \times$ less). 
On the two other data sets, \texttt{BitSet} is more than twice as fast as Roaring, but it also uses three times as much memory. When comparing the speed of \texttt{BitSet} 
and Roaring, consider that Roaring  pre-computes the cardinality at a chunk level. Thus if we need the cardinality of the aggregated bitmaps, Roaring  has the advantage.
On the \Wikileaks{} data set, Concise and WAH offer better compression than Roaring (by about \SI{30}{\percent}). This is due to the presence of long runs of ones (\texttt{11$\cdots$1} fill words), which Roaring does not compress.

Results on the 
\Censeighten{} data set are striking: Roaring is up to $900\times$ faster than the alternatives. This is due to the large differences in the cardinalities of the bitmaps. When intersecting a sparse bitmap with a dense one, Roaring is particularly efficient.

\section{Conclusion}\label{sec:conclusion}
In this paper, we introduced a new bitmap compression scheme called Roaring. It stores bitmap set entries as 32-bit integers in a space-efficient  two-level index. 
 In comparison with two competitive bitmap compression schemes, WAH and Concise, Roaring often uses less memory and is faster. 

When the data is ordered such that the bitmaps need to store long runs of consecutive values (e.g., on the \Wikileaks{} set), alternatives such as Concise or WAH may offer better compression ratios. However, even in these instances, Roaring might be faster. In future work, we will consider improving the performance and compression ratios further.
We might add new types of containers.  In particular, we might make use of fast packing techniques to optimize the storage use of the array containers~\cite{LemireBoytsov2013decoding}. We could make use of SIMD instructions in our algorithms~\cite{Polychroniou:2014:VBF:2619228.2619234,simdcompandinter, Inoue2014}. We should also review other operations beside intersections and unions, such as  threshold queries~\cite{SPE:SPE2289}. 
 
We plan to investigate further applications in information retrieval. There are reasons to be optimistic: Apache Lucene (as of version 5.0) has adopted  a Roaring format~\cite{RoaringDocIdSet} to compress document identifiers.



\bibliographystyle{wileyj}
\bibliography{bib/RoaringBitmapIPL} 

\end{document}